\documentclass[nofootinbib, superscriptaddress]{article}
\usepackage{graphicx}
\usepackage{amsmath}
\usepackage{amssymb}
\usepackage{color}
\usepackage{siunitx}
\usepackage{float}
\usepackage{makecell}
\usepackage{adjustbox}
\usepackage{tabularx}
\usepackage[
left=2.00cm,
right=2.00cm,
top=2.00cm,
bottom=2.00cm
]{geometry}
\usepackage{authblk}
\usepackage[backend=biber, style=nature, url=false, doi=false, isbn=false]{biblatex}
\DeclareUnicodeCharacter{0301}{\'{e}}

\newcommand{\vc}[1]{{\boldsymbol{\mathbf{#1}}}}

\begin{document}

\title{\textbf{A decomposition of light's spin angular momentum density}}

\author[1,2]{Alex J. Vernon}
\author[1,2]{Sebastian Golat}
\author[1]{Claire Rigouzzo}
\author[1]{Eugene A. Lim}
\author[1,2]{Francisco~J.~Rodr\'iguez-Fortu\~no\thanks{francisco.rodriguez\_fortuno@kcl.ac.uk}}

\affil[1]{Department of Physics, King's College London, Strand, London WC2R 2LS, UK}
\affil[2]{London Centre for Nanotechnology}

\date{}

\maketitle

\begin{abstract}
    Light carries intrinsic spin angular momentum (SAM) when the electric or magnetic field vector rotates over time.
    A familiar vector equation calculates the direction of light's SAM density using the right hand rule with reference to the electric and magnetic polarisation ellipses.
    Using Maxwell's equations, this vector equation can be decomposed into a sum of two distinct terms, akin to the well-known Poynting vector decomposition into orbital and spin currents.
    We present the first general study of this spin decomposition, showing that the two terms, which we call canonical and Poynting spin, are chiral analogies to the canonical and spin momenta of light in its interaction with matter.
    Both canonical and Poynting spin incorporate spatial variation of the electric and magnetic fields and are influenced by optical orbital angular momentum (OAM).
    The decomposition allows us to show that the OAM of a linearly polarised vortex beam can impart a first-order preferential force to chiral matter in the absence of spin.
    
\end{abstract}

\section{Introduction}

Optical angular momentum is contributed to by physical orbital and spin parts \cite{Barnett2010,Barnett2016,VanEnk1994_2,Bliokh2015}.
The orbital component can be either extrinsic or intrinsic and is associated with the spatial structure of the light field.
Orbital angular momentum (OAM) is most famously stirred into light by optical vortices, which have helical phase fronts twisted around a one-dimensional line singularity \cite{Allen1992,Yao2011,Dennis2009,Berry2000}.
Spin angular momentum (SAM), meanwhile, develops as the electric and magnetic field vectors rotate during oscillation \cite{Cameron2012}, and is an intrinsic quantity, independent of co-ordinate origin.
Both OAM and SAM arise from chiral structures in light, and as such, can access the chirality of matter, enhancing, pushing, twisting, and torquing with or without a parity bias \cite{CanaguierDurand2013,Wang2014}.
Many commonly observed chiroptical interactions, such as circular dichroism, arise from the coupling of chiral matter to a photon's spin state.
Recent attention has also turned to OAM-dependent photon absorption and scattering, with a range of effects theorised \cite{Forbes2021,Forbes2019,Ye2019,Forbes2019_2}.
Chiroptical effects involving OAM often hinge on the strong longitudinal field component introduced by beam focussing \cite{Forbes2021_2,Forbes2022}.
Unlike for photons with opposite spins, however, preferential absorption of photons with different OAM handedness does not occur in the dipole approximation for paraxial light \cite{Andrews2004,Araoka2005}.

A well-known orbital and spin separation can be performed locally in the Poynting vector \cite{Berry2009} (kinetic momentum density, when divided by $c^2$).
The Poynting vector may be expressed as a sum of the orbital current, pointing in the direction of canonical momentum, and spin current, proportional to the virtual spin momentum.
This is a physically meaningful decomposition, tying in to the famous Abraham-Minkowski dilemma \cite{Barnett2010_2,Bliokh2017}.

Surprisingly to most readers, a similar vector decomposition can be performed on the electromagnetic spin density $\mathbf{S}$, which is split into, perhaps confusingly, orbital-like and spin-like contributions to the total electromagnetic spin.
We name these two terms canonical and Poynting spin.
This little-known decomposition is the focus of our work---amidst discussion by Shi et al who, without giving the complete decomposition explicitly, linked the two terms to longitudinal and transverse spin in some specific cases \cite{Shi2021,Shi2022,Shi2023}, no clear picture exists in the literature on the physical significance of the two terms in general, in the same way as for the decomposed Poynting vector.
The decomposition as we present has been expressed previously (e.g. \cite{Bliokh2014}), but with almost no discussion on the function of terms as component parts of $\mathbf{S}$.

We therefore have two objectives: to clarify and understand the meaning of the orbit-like and spin-like components of $\mathbf{S}$, and to emphasise their value for the research community in interpreting electromagnetic SAM density.
For example, by use of the Maxwell stress tensor, one understands that the canonical and Poynting spin vectors are responsible for different interactions between light and chiral particles.
While the two decomposed spin vectors are organised independently in general monochromatic light, a remarkable implication lies in beams carrying OAM, such as vortex beams.
OAM imparts a longitudinal component to both canonical and Poynting spins, even in the absence of longitudinal total spin.
By applying the decomposition to a linearly polarised vortex beam, we show that a total spin-free longitudinal chiral force exists---a force which can act even under the dipole approximation, despite originating from the beam's OAM.

We generalise the spin decomposition to time-dependent fields and four-vector representation, unlocking a deeper understanding of the behaviour of the two terms. We also show that the ability to split spin into two terms appears to be a general feature of wave fields, similarly to the decomposition of the Poynting vector, including those of the linearised theories of gravity \cite{Barnett2014} and acoustics, which have distinctly different vector structures.
Throughout, we use bold latin letters $\mathbf{E}$ and $\mathbf{H}$ to represent time-harmonic complex electromagnetic field phasors, which are a function of position only [e.g. $\mathbf{E(\mathbf{r})}=\mathbf{E}_0\exp(i\mathbf{k\cdot r})$], and scripted characters $\boldsymbol{\mathcal{E}}$ and $\boldsymbol{\mathcal{H}}$ denoting real, time-varying field vectors (e.g. $\boldsymbol{\mathcal{E}}(\mathbf{r},t)=\textrm{Re}\{\mathbf{E(\mathbf{r})}\exp(-i\omega t)\}$).

\begin{figure}[t]
    \centering
    \includegraphics[width=0.5\textwidth]{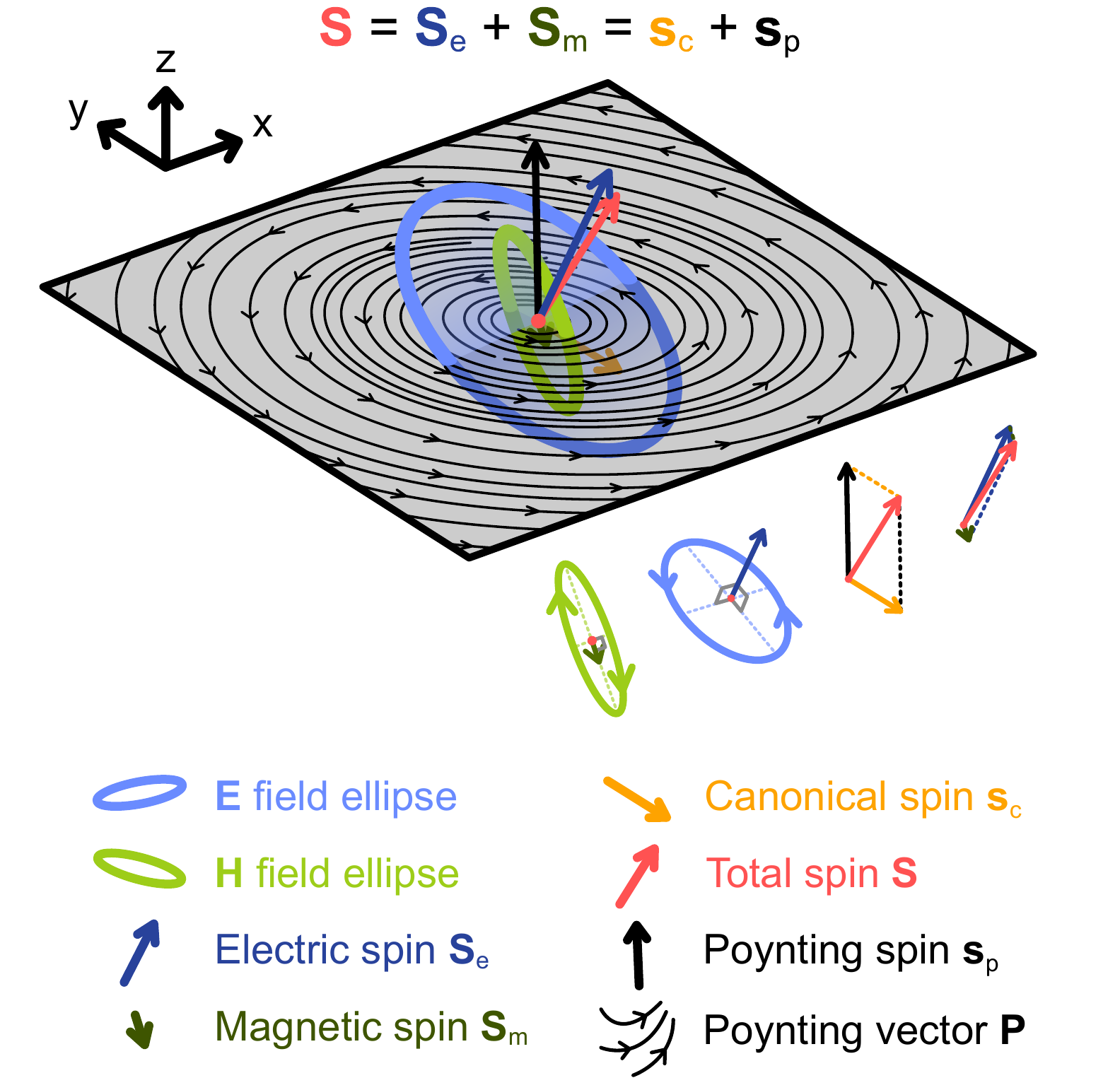}
    \caption{
    The component parts of light's total spin angular momentum density, visualised at a single point (red circle) in the 3D interference of monochromatic plane waves.
    Enlarged electric and magnetic field ellipses and coloured vector arrows, representing the total spin $\mathbf{S}$, electric and magnetic spins $\mathbf{S}_{e}$ and $\mathbf{S}_{m}$, and the canonical $\mathbf{s}_c$ and Poynting spins $\mathbf{s}_p$ of the decomposition Eq. \eqref{spin_decomp}, are plotted.
    The electric and magnetic spin vectors are normal to the $\mathbf{E}$ and $\mathbf{H}$ polarisation ellipses, according to the right-hand rule and the sense of rotation of the instantaneous field vectors (see separated diagrams below the main combined image).
    Projected on the $xy$ plane are the Poynting vector streamlines, which in this case organise a clear curling structure.
    This provides a visual aid for Poynting spin (large black arrow), which is informed by three-dimensional curling of the Poynting vector.
    In structured light where the $\mathbf{E}$ and $\mathbf{H}$ ellipses are de-coupled, the canonical spin and therefore chiral pressure (Eq. \eqref{eq:chiral_force}) can point in a completely different direction to the any of the total spin or its electric and magnetic contributions.
    }\label{fig_visual_spin}
\end{figure}

\section{Electromagnetic spin decomposition}

We will first contextualise the decomposition of electromagnetic SAM density by recalling a well-known orbital and spin decomposition which exists in free-space and time-harmonic fields for the time-averaged Poynting vector,
\begin{equation}\label{Poyntingvector}
\mathbf{P}=\frac{1}{2}\textrm{Re}\{\mathbf{E^*\times H}\},
\end{equation}
which represents the flux of active power in the light field.
The Poynting vector is separated into the orbital and spin currents \cite{Berry2009,Bliokh2017},
\begin{equation}\label{poynting_decomp}
    \mathbf{P}=\underbrace{\frac{c^2}{4\omega}\textrm{Im}\{\epsilon_0\mathbf{E^*}\cdot(\nabla)\mathbf{E}+\mu_0\mathbf{H^*}\cdot(\nabla)\mathbf{H}\}}_{\mathbf{p}_o}+\underbrace{\frac{c^2}{2}\nabla\times\overbrace{\frac{1}{4\omega}\textrm{Im}\{\epsilon_0\mathbf{E^*}\times\mathbf{E}+\mu_0\mathbf{H^*}\times\mathbf{H}\}}^{\mathbf{S}}}_{\mathbf{p}_s},
\end{equation}
defined using the complex phasors $\mathbf{E}$ and $\mathbf{H}$.
The inner product notation is $\mathbf{a}\cdot(\nabla)\mathbf{b}=a_x\nabla b_x+a_y\nabla b_y+a_z\nabla b_z$.
When divided by $c^2$, $\mathbf{P}$ has units of a momentum density and is termed the kinetic momentum density, while the orbital current $\mathbf{p}_o$ becomes the canonical momentum density, corresponding to the expectation value of linear momentum carried by photons at each point in space (unlike kinetic momentum, the canonical momentum is a directly measurable quantity).
The net flow of the instantaneous Poynting vector into or out of a volume relates to the change in electromagnetic energy density over time, according to a continuity equation.
An adjacent conserved quantity of light is its chirality, which has its own continuity equation and an associated chirality flux \cite{Bliokh2011,Tang2010,Lipkin1964}.
Chirality $\chi$ is one of an infinite hierarchy of conserved quantities in linear, non-dispersive media, to which helicity $h$ also belongs \cite{Lipkin1964,Cameron2012}---helicity and chirality are often conflated because they (and their fluxes) are proportional by a factor of $\omega^2$ in monochromatic waves, though in general they are distinct.
In monochromatic light the flux of helicity corresponds to the time-averaged SAM density (often referred to simply as `spin'), given by,
\begin{equation}\label{def_spin}
    \mathbf{S}=\frac{1}{4\omega}\textrm{Im}\{\epsilon_0\mathbf{E^*\times E}+\mu_0\mathbf{H^*\times H}\} \equiv \mathbf{S}_{e}+ \mathbf{S}_{m} \ .
\end{equation}
Spin's individual electric and magnetic contributions point in the normal direction to the electric and magnetic polarisation ellipses, drawn over time by the instantaneous vectors $\boldsymbol{\mathcal{E}}(\mathbf{r},t)$ and $\boldsymbol{\mathcal{H}}(\mathbf{r},t)$.
Like the Poynting vector, and using the same procedure, spin $\mathbf{S}$ may be split into a sum of two vectors: the canonical spin $\mathbf{s}_c$ and the Poynting spin $\mathbf{s}_p$.
This decomposition is our main focus and is, explicitly,
\begin{equation}\label{spin_decomp}
    \mathbf{S}=\underbrace{\frac{1}{4\omega^2}\textrm{Re}\{\mathbf{E^*}\cdot(\nabla)\mathbf{H}-\mathbf{H^*}\cdot(\nabla)\mathbf{E}\}}_{\mathbf{s}_c}+\underbrace{\frac{1}{2\omega^2}\nabla\times\overbrace{\frac{1}{2}\textrm{Re}\{\mathbf{E^*\times H}\}}^\mathbf{P}}_{\mathbf{s}_p}.
\end{equation}
Any subsequent mention of `canonical spin' in this work always refers to the first term, $\mathbf{s}_c=\frac{1}{4\omega^2}\textrm{Re}\{\mathbf{E^*}\cdot(\nabla)\mathbf{H}-\mathbf{H^*}\cdot(\nabla)\mathbf{E}\}$, and `Poynting spin' to the second, $\mathbf{s}_p=\frac{1}{2\omega^2}\nabla\times\mathbf{P}$.
Their sum $\mathbf{S}$ (Eq. (\ref{def_spin})) we refer to as total spin.
This naming scheme reflects that of the decomposed kinetic momentum, split into canonical momentum and spin momentum.

\section{Physical interpretation}

Previous works \cite{Shi2021,Shi2022} have linked the canonical $\mathbf{s}_c$ and Poynting spin $\mathbf{s}_p$ vectors to the longitudinal and transverse spin of light, respectively. However, this is only true in a very limited number of cases such as linearly polarised evanescent waves. It cannot be a general result because well-defined longitudinal and transverse directions do not exist for many-plane wave or multiple beam interference. In fact, it does not hold even when there is a well-defined longitudinal direction: a circularly polarised evanescent wave has both longitudinal and transverse spin components contained in the Poynting spin vector.

One way to gain an understanding of the physical significance of the canonical and Poynting spins is to study the interactions between light and chiral matter.
Chiral light can exert preferential forces on enantiomers with opposite handedness, with direct proportionality to the light's SAM.
When light shines on a particle much smaller than the wavelength (Rayleigh regime) it gets polarised and acquires an electric $\mathbf{p}$ and a magnetic $\mathbf{m}$ dipole.
In the linear regime, the dipole moments are proportional to the incident fields, $\mathbf{p}=\alpha_\text{e}\varepsilon\mathbf{E}+i\alpha_\text{c}\mathbf{H}/c$ and $\mathbf{m}=\alpha_\text{m}\mathbf{H}-i\alpha_\text{c}\mathbf{E}/\eta$, where $\alpha_\text{e}$, $\alpha_\text{m}$ and $\alpha_\text{c}$ are the electric, magnetic, and chiral polarisabilities of the particle.
The chiral polarisability $\alpha_\text{c}$ is a pseudoscalar so it changes sign between the two enantiomers (mirror-reflected versions), and vanishes unless the matter is chiral.
Under these assumptions, the illuminating light exerts a chiral optical force on the particle (which changes sign for the different enantiomers) given by \cite{Yoo2019,Hayat2015,Zhang2017,Cao2018,NietoVesperinas2010}:

\begin{equation}
\label{eq:chiral_force}
{\mathbf{F}_\text{chiral}}=\underbrace{\omega\nabla[\mathrm{Re}(\alpha_\text{c}) h]}_\text{helicity gradient}+\underbrace{2\omega k[\mathrm{Im}(\alpha_\text{c})\mathbf{s}_\text{c}]}_\text{chiral pressure}-\underbrace{\,\omega \frac{k^4}{3\pi}[\mathrm{Re}(\alpha_\text{e}^\ast\alpha_\text{c}) \mathbf{S}_\text{e}\!+\!\mathrm{Re}(\alpha_\text{m}^\ast\alpha_\text{c})\mathbf{S}_\text{m}]}_\text{spin recoil},
\end{equation}

\noindent where $k=\omega/c$ is the wavenumber, $h=-\mathrm{Im}(\mathbf{E}^*\cdot\mathbf{H})/(2\omega c)$ is the cycle-averaged optical helicity density \cite{Forbes2022, Bliokh2013}, and $\mathbf{S}_e$ and $\mathbf{S}_m$ are the electric and magnetic parts of the total spin $\mathbf{S}$ from Eq.~(\ref{def_spin}).
One immediately sees that the canonical spin $\mathbf{s}_\text{c}$, one of the two terms in the spin decomposition [Eq.~(\ref{spin_decomp})], appears in this force equation and is directly responsible for the chiral pressure (in this context, the product $k\mathbf{s}_\text{c}$ is sometimes called the chiral momentum \cite{Bliokh2014}), while the \emph{total} spin appears in the spin recoil term.
Poynting spin contributes (together with canoncial spin) only to the higher-order recoil force caused by the unbalanced radiation pattern of an electric-magnetic dipole.
This also implies that in an electromagnetic field whose SAM is pure Poynting spin (i.e., zero canonical spin), the particle will experience no chiral pressure, only helicity gradients and relatively weaker spin recoil terms of the force.

An alternative route to the physical meaning of the canonical and Poynting spins is found purely in the two terms' mathematical expressions as shown in Eq.~(\ref{spin_decomp}).
The Poynting spin, being proportional to the curl of the Poynting vector, is easily interpreted as the vorticity in energy flow.
The expression for the canonical spin is much harder to interpret initially, but begins to unravel if we decompose a general electromagnetic field into circularly polarised plane waves.
Any arbitrary electromagnetic field, no matter how complicated, can be expressed in momentum space by an angular spectrum (an infinite sum of plane waves of different wavevectors, weighted by an amplitude function).
Using this property, we can further probe the canonical spin term in monochromatic light by separating electric and magnetic fields into two component fields of opposite helicities, indicated by the $+$ and $-$ subscripts,
\begin{equation}\label{angspectrumE}
    \mathbf{E}(\mathbf{r})=\!\!\iiint\tilde{\mathbf{E}}(\mathbf{k})e^{i\mathbf{k}\cdot\mathbf{r}}\mathrm{d}^3k=\!\!\iiint[\tilde{{E}}_+(\mathbf{k})\hat{\mathbf{e}}_+(\mathbf{k})+\tilde{{E}}_-(\mathbf{k})\hat{\mathbf{e}}_-(\mathbf{k})]e^{i\mathbf{k}\cdot\mathbf{r}}\mathrm{d}^3k=\mathbf{E}_+(\mathbf{r})+\mathbf{E}_-(\mathbf{r}),
\end{equation}
where $\hat{\mathbf{e}}_\pm(\mathbf{k})$ represent the circularly polarised unit vectors for each plane wave with wave-vector $\mathbf{k}$ (see, e.g., \cite{Wei2020}). The magnetic field's associated helicity components are obtained from Faraday's law $\nabla\times{\mathbf{E}}=i\omega\mu\mathbf{H}$ as,
\begin{equation}\label{angspectrumH}
    \mathbf{H}(\mathbf{r})=\frac{1}{\eta}\!\iiint\frac{\mathbf{k}}{k}\times\tilde{\mathbf{E}}(\mathbf{k})e^{i\mathbf{k}\cdot\mathbf{r}}\mathrm{d}^3k=\frac{1}{\eta}\!\iiint[-i\tilde{{E}}_+\hat{\mathbf{e}}_+(\mathbf{k})+i\tilde{{E}}_-\hat{\mathbf{e}}_-(\mathbf{k})]e^{i\mathbf{k}\cdot\mathbf{r}}\mathrm{d}^3k=\mathbf{H}_+(\mathbf{r})+\mathbf{H}_-(\mathbf{r}).
\end{equation}
where we used the property $(\mathbf{k}/k)\times\hat{\mathbf{e}}_\pm = \mp i \hat{\mathbf{e}}_\pm$.
Equations (\ref{angspectrumE}) and (\ref{angspectrumH}) show that the helicity-separated electric and magnetic fields are related by $\mathbf{H}_\pm(\mathbf{r})=\mp i\mathbf{E}_\pm(\mathbf{r})/\eta$, characteristic of a circularly polarised plane wave---this is true not only in the spectral representation, but also in the spatial representation, for any arbitrary field (a property explored in depth in \cite{FernandezThesis}).
This allows us to substitute helicity-separated fields into many dual quantities, including canonical momentum and canonical spin, gaining further insight.
Simply substituting $\mathbf{E}=\mathbf{E}_++\mathbf{E}_-$ and $\mathbf{H}=-i\left(\mathbf{E}_+-\mathbf{E}_-\right)/\eta$ into the expression for orbital current [$\mathbf{p}_o$ of Eq. (\ref{poynting_decomp})], we find, after some algebra,
\begin{equation}
    \mathbf{p}_o=\frac{c^2}{2\omega}\epsilon_0\textrm{Im}\{\mathbf{E}^*_+\cdot(\nabla)\mathbf{E}_++\mathbf{E}^*_-\cdot(\nabla)\mathbf{E}_-\}=c^2(\mathbf{p}_++\mathbf{p}_-),
\end{equation}
showing that the helicity segregation of the $\mathbf{E}$ and $\mathbf{H}$ fields translates to a separation of contributions to the field's momentum $\mathbf{p}=\mathbf{p}_++\mathbf{p}_-$ by photons of positive and negative helicity.
Note that orbital current is proportional to canonical momentum by $\mathbf{p}_o=c^2\mathbf{p}$; here, $\mathbf{p}_+$ and $\mathbf{p}_-$ are helicity-separated momentum densities.
Making the same substitution in the expression for canonical spin $\mathbf{s}_c$ [first term of Eq. (\ref{spin_decomp})] illuminates its meaning,
\begin{equation}
\begin{split}
    \mathbf{s}_c&=\frac{1}{4\omega^2}\textrm{Re}\big\{\!-\frac{i}{\eta}[\mathbf{E^*_+}+\mathbf{E^*_-}]\cdot(\nabla)[\mathbf{E_+}-\mathbf{E_-}]+\frac{i}{\eta}[\mathbf{E^*_+}-\mathbf{E^*_-}]\cdot(\nabla)[\mathbf{E_+}+\mathbf{E_-}]\big\}\\&=\frac{1}{2k\omega}\epsilon_0\textrm{Im}\{\mathbf{E}^*_+\cdot(\nabla)\mathbf{E}_+-\mathbf{E}^*_-\cdot(\nabla)\mathbf{E}_-\}=\frac{1}{k}\left(\mathbf{p}_+-\mathbf{p}_-\right).
\end{split}
\end{equation}
Canonical spin is proportional to the difference in linear momentum densities carried by photons of oppositely signed helicity.
This brings a clear physical interpretation of the canonical spin, and also helps explain why the chiral pressure force acts in the direction of $\mathbf{s}_c$, as photons of opposite helicities are absorbed or scattered in different amounts by chiral particles.
In a general structured field, $\mathbf{p}_+-\mathbf{p}_-\neq\mathbf{p}_++\mathbf{p}_-$ and hence canonical spin is decoupled from any local longitudinal direction defined by canonical momentum.

\section{Mathematical derivation}
In this section, we briefly lay out the mathematical steps taken to arrive at Eq.~(\ref{spin_decomp}), before developing the expression into a more fundamental 4-vector description.
For each of the canonical and Poynting spin terms, the 4-vector decomposition incorporates a time component which characterises how the two terms transform differently between reference frames.
\subsection{3-vector decomposition}
Maxwell’s equations enable both decompositions of the Poynting vector Eq.~(\ref{poynting_decomp}) and spin Eq.~(\ref{spin_decomp}) by bringing the relevant vector from its usual representation into a form which can be separated into two terms using a vector identity,
\begin{equation}\label{identity}
    \mathbf{a}\times(\nabla\times \mathbf{b})=\mathbf{a}\cdot(\nabla)\mathbf{b}-(\mathbf{a}\cdot\nabla)\mathbf{b},
\end{equation}
where $\mathbf{a}$ and $\mathbf{b}$ are arbitrary vectors, and the previously unseen notation on the right hand side means $(\mathbf{a}\cdot\nabla)\mathbf{b}=a_x \partial_x\mathbf{b}+a_y \partial_y\mathbf{b}+a_z \partial_z\mathbf{b}$.
In time-harmonic fields, the phasors $\mathbf{E}$ and $\mathbf{H}$ can be substituted for curls of their counterpart via Faraday's and Ampere's laws.
Substituting for the un-conjugated phasors in the spin vector definition Eq. (\ref{def_spin}) gives an expression in the form of Eq. (\ref{identity}),
\begin{equation}
    \mathbf{S}=\frac{1}{4\omega}\textrm{Im}\big\{\epsilon_0\mathbf{E^*}\times\Big(\frac{i}{\omega\epsilon_0}\nabla\times\mathbf{H}\Big)-\mu_0\mathbf{H^*}\times\Big(\frac{i}{\omega\mu_0}\nabla\times\mathbf{E}\Big)\big\}.
\end{equation}
Applying the vector identity Eq. (\ref{identity}), we have,
\begin{equation}\label{decomp_step2}
    \mathbf{S}=\frac{1}{4\omega^2}\textrm{Re}\{\mathbf{E^*}\cdot(\nabla)\mathbf{H}-\mathbf{H^*}\cdot(\nabla)\mathbf{E}\}+\frac{1}{4\omega^2}\textrm{Re}\{(\mathbf{H^*}\cdot\nabla)\mathbf{E}-(\mathbf{E^*}\cdot\nabla)\mathbf{H}\}.
\end{equation}
Gauss' law in free space ($\nabla\cdot\mathbf{E}=0$ and $\nabla\cdot\mathbf{H}=0$), combined with a second vector identity $\nabla\times(\mathbf{a}\times\mathbf{b})=\mathbf{a}(\nabla\cdot\mathbf{b})-\mathbf{b}(\nabla\cdot\mathbf{a})+(\mathbf{b}\cdot\nabla)\mathbf{a}-(\mathbf{a}\cdot\nabla)\mathbf{b}$, extracts the curl of the Poynting vector from the second $\textrm{Re}\{\}$ term of Eq. (\ref{decomp_step2}),
\begin{equation}
    \frac{1}{4\omega^2}\textrm{Re}\{(\mathbf{H^*}\cdot\nabla)\mathbf{E}-(\mathbf{E^*}\cdot\nabla)\mathbf{H}\}=\frac{1}{2\omega^2}\nabla\times\frac{1}{2}\textrm{Re}\{\mathbf{E^*\times H}\}=\mathbf{s}_p,
\end{equation}
completing the final step in obtaining the electromagnetic spin decomposition as we present in Eq. (\ref{spin_decomp}).

In addition to the monochromatic case, canonical and Poynting spin analogies for the flow of chirality may also be defined in polychromatic fields, expressed using the full time-dependent fields $\boldsymbol{\mathcal{E}}(\mathbf{r}, t)$ and $\boldsymbol{\mathcal{H}}(\mathbf{r}, t)$ and stemming from the instantaneous flow of chirality $\boldsymbol{\mathcal{F}}(\mathbf{r}, t)$ \cite{Bliokh2011,Tang2010,Vazquez2018,Lipkin1964}.
\begin{equation}\label{chi_flow}
    \boldsymbol{\mathcal{F}}=\frac{1}{2}\left(\boldsymbol{\mathcal{E}}\times(\nabla\times\boldsymbol{\mathcal{H}})-\boldsymbol{\mathcal{H}}\times(\nabla\times\boldsymbol{\mathcal{E}})\right).
\end{equation}
Chiral flow $\boldsymbol{\mathcal{F}}$, unlike the flux of helicity in time-dependent fields, is defined without the vector potentials $\boldsymbol{\mathcal{A}}$ and $\boldsymbol{\mathcal{C}}$, which avoids gauge complications and contains a direct relation to the instantaneous Poynting vector $\boldsymbol{\mathcal{P=E\times H}}$.
For monochromatic light, $\boldsymbol{\mathcal{F}}$ is not time-dependent and becomes proportional to the field's SAM density $\mathbf{S}$.
Equation (\ref{chi_flow}) is immediately presented in a form which can be broken into two vectors using the two identities detailed in this section, giving the extension of the time-harmonic spin decomposition to time-dependent fields,
\begin{equation}\label{F_decomp}
    \boldsymbol{\mathcal{F}}=\frac{1}{2}\left[\left(\boldsymbol{\mathcal{E}}\cdot(\nabla)\boldsymbol{\mathcal{H}}-\boldsymbol{\mathcal{H}}\cdot(\nabla)\boldsymbol{\mathcal{E}}\right)+\nabla\times(\boldsymbol{\mathcal{E\times H}})\right].
\end{equation}
The above is a more general decomposition of chirality flow than Eq.~(\ref{spin_decomp}), and is valid for polychromatic aperiodic light, at every instant in time.
It is worth stressing again that the flow of chirality $\boldsymbol{\mathcal{F}}$ is a different quantity to the flow of helicity (SAM density) for general time-dependent fields.
To obtain the helicity flow equivalent of Eq. (\ref{F_decomp}) in the Coulomb gauge, one simply substitutes $\boldsymbol{\mathcal{E}}\rightarrow\boldsymbol{\mathcal{A}}$ and $\boldsymbol{\mathcal{H}}\rightarrow\boldsymbol{\mathcal{C}}$.
Equation \eqref{F_decomp} assumed a source-free medium, such that $\nabla \cdot \boldsymbol{\mathcal{E}} = \rho/\epsilon = 0$. Dropping this assumption, Eq.~\eqref{F_decomp} can be further generalised by adding a third term, $\frac{1}{2}\left[\frac{\rho}{\epsilon}\boldsymbol{\mathcal{H}}\right]$, to the decomposition.

\subsection{4-vector representation}\label{sec:fourvector}
%4-vector notation was introduced by Minkowski \cite{Minkowski} as well as Einstein \cite{Einstein}, and is particularly elegant and useful to verify the Lorentz covariance of quantities and equations.
In an attempt to extract new physical insight, we reproduce the above spin decomposition Eq.~\eqref{spin_decomp} in 4-vector notation.
This notation is particularly elegant and useful to verify the Lorentz covariance of quantities and equations.
We will begin by introducing some common definitions.
Electrodynamics is intrinsically relativistic, and several related physical quantities can be expressed via well-known 4-vectors, such as the 4-potential (grouping the scalar and vector potentials) and the 4-current (grouping the charge and current densities).
4-vector notation is well known to provide an efficient way to formulate Maxwell's equations.
Indeed, the time-dependent Maxwell's equations (using scripted vectors) in vacuum are a combination of four equations, two scalar equations $\mathbf{\nabla}\cdot \boldsymbol{\mathcal{E}}=0, \mathbf{\nabla}\cdot \boldsymbol{\mathcal{H}}=0$ and two vector equations $\mathbf{\nabla} \times \boldsymbol{\mathcal{E}}=- \mu_0\frac{\partial \boldsymbol{\mathcal{H}}}{ \partial t} , \mathbf{\nabla} \times \boldsymbol{\mathcal{H}}= \epsilon_0 \frac{\partial \boldsymbol{\mathcal{E}}}{\partial t}$.
In 4-vector notation, however, we can write Maxwell's equations as: 
\begin{equation}
    \partial_\alpha \mathcal{F}^{\alpha \beta}=0, \qquad \epsilon^{\alpha \beta \gamma\lambda}\partial_\alpha \mathcal{F}_{\beta \gamma} =0 .
    \label{Maxwell_eq_tensorial}
\end{equation}
Throughout, the greek indices $\mu,\nu,\alpha,\beta$... run from $0$ to $3$, where the $0^{th}$ component labels the time direction.
Repeated indices are summed over.
Note that the distinction between subscript/superscript placement of indices are important, as $A_{\mu} = \eta_{\mu\nu}A^{\nu}$ where the so-called Minkowski metric $\eta_{\mu\nu} = \mathrm{diag}(-1,1,1,1)$.
We also use Roman indices $i=1,2,3$ to denote the three-dimensional space.
In Eq.~(\ref{Maxwell_eq_tensorial}), $\epsilon^{\alpha \beta \gamma\lambda}$ is the Levi-Civita symbol \cite{tyldesley1975introduction}, while $\mathcal{F}_{\alpha \beta}$ is the field strength tensor that conveniently packages the electric and magnetic fields:
\begin{equation}
\mathcal{F}_{\mu \nu}=\begin{pmatrix}
0 &- \mathcal{E}_x/c & -\mathcal{E}_y/c  & -\mathcal{E}_z/c \\
\mathcal{E}_x/c & 0 & -\mu_0 \mathcal{H}_z & \mu_0 \mathcal{H}_y \\
\mathcal{E}_y/c  & \mu_0 \mathcal{H}_z & 0 & -\mu_0 \mathcal{H}_x \\
\mathcal{E}_z/c &  -\mu_0 \mathcal{H}_y & \mu_0 \mathcal{H}_x & 0
\end{pmatrix}.
\end{equation}
The field strength tensor can be expressed as $\mathcal{F}_{\alpha \beta}=\partial_\alpha \mathcal{A}_\beta - \partial_\beta \mathcal{A}_\alpha$, where $\mathcal{A}_\alpha$ is the 4-vector potential, that encases the scalar potential $\varphi$ and the vector potential $\boldsymbol{\mathcal{A}}$ and $\partial_\mu$ is a generalisation of gradient:
\begin{equation}
\mathcal{A}^\mu=\begin{pmatrix}
\mathcal{A}^0 \\
\mathcal{A}^i
    \end{pmatrix}=
    \begin{pmatrix}
    \varphi /c \\
\boldsymbol{\mathcal{A}}
    \end{pmatrix},
    \label{vector_potential}
\quad
\partial_\mu=\begin{pmatrix}
\frac{\partial}{c\partial t} \\
\partial_i
    \end{pmatrix}=
    \begin{pmatrix}
    \frac{\partial}{c\partial t} \\
\nabla
    \end{pmatrix}.
\end{equation} 
Equation (\ref{Maxwell_eq_tensorial}) is a striking example of the neatness of the tensorial notation: the physical content is the same, but the tensorial notation is clearer and exhibits Lorentz covariance.
For time-harmonic fields, these tensors will also have a phasor representation (e.g. ${\mathcal{F}}_{\mu\nu}(\mathbf{r},t)=\textrm{Re}\{{F_{\mu\nu}(\mathbf{r})}\exp(-i\omega t)\}$), including the time components of the gradient, which for phasors becomes $\partial_0=-i\omega/c$.

We now have the tools to write, using phasors, the time-averaged spin decomposition Eq. \eqref{spin_decomp} in tensorial notation.
The first step is to equivalently express the spin, defined in Eq.~\eqref{def_spin}, using spatial indices to replicate the vector operations,
\begin{equation}
    S^i=\frac{1}{4\omega}\epsilon^{ijk}\textrm{Im}\{\epsilon_0 E^*_j  E_k+\mu_0 H^*_j H_k\}.
\end{equation}
Then, the equivalent of the proposed decomposition in Eq.~\eqref{spin_decomp} is: 
\begin{equation}
    S^i=\underbrace{\frac{1}{4\omega^2}\textrm{Re}\{E^*_j \partial_i H^j-H^*_j \partial_i E^j\}}_{s^i_c}+\underbrace{\frac{1}{2\omega^2} \epsilon^{ijk}\partial_j \frac{1}{2}\textrm{Re}\{\epsilon_{klm} E^{*l} 
    H^m\}}_{s^i_p}.\label{spin_decomp_3D}\end{equation}
The next step is to find a 4-vector, for which the spatial part would reduce to Eq.~\eqref{spin_decomp_3D}.
% To do so, we need to express Eq. \eqref{spin_decomp_3D} using only tensorial quantities like $A^\mu$ and $F_{\mu \nu}$.
It turns out that this 4-vector quantity is exactly the helicity density and flux, a 4-current density associated with conserved helicity \cite{Cameron_2012}
\begin{equation}\label{eq:heli}
    S^\mu= \frac{1}{4} \operatorname{Re}\left\{A_\nu^* G^{\nu \mu}+ C_\nu^* F^{\nu \mu}\right\},
\end{equation}
where $C^{\mu} = (\psi ,-i\omega\mu_0^{-1}(\mathbf{\nabla} \times \mathbf{A}))^\intercal $ is the magnetic equivalent to the 4-potential $A_\mu$, and $G_{\mu \nu}=\partial_\nu C_\mu-\partial_\mu C_\nu$ is the corresponding field strength tensor. Note that Eq.~\eqref{eq:heli} is time-averaged and the fields are in their phasor representation.
We include the instantaneous variant in the supplementary information.
Working in the Coulomb gauge, i.e., the scalar and vector potentials are chosen such that $\nabla\cdot\mathbf{A}=\phi=0$ and $\nabla\cdot\mathbf{C}=\psi=0$, and after some algebra (see supplementary Mathematica file \cite{Golat_Rigouzzo}), we find that the spatial part of the 4-vector $S^\mu$ reproduces Eq. \eqref{spin_decomp_3D}, whilst the temporal part is the cycle-averaged helicity density $h=-\textrm{Im}\{\mathbf{E^*\cdot H}\}/(2\omega c)$,
\begin{equation}\label{eq:spin_density_vector}
    S^\mu=\frac{1}{4 \omega } \operatorname{Im}\left\{\left(\begin{array}{c}
 -\frac{2}{c}\mathbf{E}^* \cdot \mathbf{H} \\
 \epsilon_0\mathbf{E}^*\times \mathbf{E}+\mu_0 \mathbf{H}^* \times \mathbf{H}
 \end{array}\right)\right\}=\left(\begin{array}{c}
  h \\
 \mathbf{S}
 \end{array}\right) ~.
 \end{equation}
This is consistent with the fact that $h$ and $\mathbf{S}$ are density and flux associated with integrated helicity, the conserved quantity associated with the dual symmetry \cite{Cameron2012,Lipkin1964} (hence $\nabla\cdot\mathbf{S}=0$ for free-space monochromatic fields).
This four-vector current $S^\mu$ has a decomposition equivalent to Eq. \eqref{spin_decomp},
\begin{equation}
    S^{\mu} =S^{\mu}_C +S^{\mu}_P,
\end{equation}
where the individual components are  given by,
\begin{equation}
\begin{aligned}
&  S_{C}^\mu=\frac{1}{4} \operatorname{Re}\left\{A_\nu^*\left(\partial^\mu C^\nu\right)-C_\nu^*\left(\partial^\mu A^\nu\right)\right\}=\frac{1}{4 \omega^2}\left(\begin{array}{c}
-2(\omega/c) \operatorname{Im}\left\{\mathbf{E}^*\cdot\mathbf{H}\right\} \\
\operatorname{Re}\left\{\mathbf{E}^* \cdot(\mathbf{\nabla}) \mathbf{H}-\mathbf{H}^* \cdot(\mathbf{\nabla}) \mathbf{E}\right\}
\end{array}\right)=\left(\begin{array}{c}
h \\
\mathbf{s}_{c}
\end{array}\right), \\
&  S_{P}^\mu=\frac{1}{4} \operatorname{Re}\left\{C_\nu^*\left(\partial^\nu A^\mu\right)-A_\nu^*\left(\partial^\nu C^\mu\right)\right\}=\frac{1}{4 \omega^2}\left(\begin{array}{c}
0 \\
 \operatorname{Re}\left\{\left(\mathbf{H}^* \cdot \nabla\right) \mathbf{E}-\left(\mathbf{E}^* \cdot \mathbf{\nabla}\right) \mathbf{H}\right\}
\end{array}\right)=\left(\begin{array}{c}
0 \\
\mathbf{s}_{p}
\end{array}\right),\label{poynting_canonical_spin}
\end{aligned}
\end{equation}
reproducing the spin decomposition in 4-vector notation.
Note that the temporal part of the total spin, the helicity density, is carried only by the canonical 4-spin term while the time component of the Poynting 4-spin is zero.

Let us briefly comment on the choice of the Coulomb gauge.
As can be seen from the supplementary information, the decomposition of $S^\mu$ is independent of the choice of frame and gauge.
As already discussed in previous work (e.g. \cite{Cameron2012,Cameron_2012}), while the local density $S^\mu$ is not gauge independent, upon integration one gets \textit{integrated helicity}, which is a gauge-independent quantity that is conserved in any scenario where there is electromagnetic dual symmetry.
Furthermore, the integral depends only on the transverse components of the potentials, which makes the Coulomb gauge the most convenient gauge to work with, and in this gauge, the helicity flux (spatial component of $S^\mu$) also coincides with the optical spin density.
Note, however, that the gauge fixed version of this decomposition is not invariant under Lorentz boosts, unless we simultaneously change our potentials to the Coulomb gauge associated with the new boosted frame. 

\section{Spin decomposition examples}
It is instructive to apply the spin decomposition in Eq.~(\ref{spin_decomp}) to some simple examples of light: a general evanescent wave and three commonly known focused beams, namely a Gaussian beam, a radial/azimuthal beam ($l=0$), and an $l=1$ vortex beam (all linearly polarised in the transverse plane).
What differentiates each of these fields are their energy density and phase structures, due to one-dimensionally monotonic, doughnut or Gaussian real-space amplitude profiles, and the presence of OAM, two characteristics which reorient the Poynting vector throughout space and generate different amounts of Poynting spin.
Poynting spin can emerge even when the instantaneous vectors $\boldsymbol{\mathcal{E}}$ and $\boldsymbol{\mathcal{H}}$ do not actually rotate, because its counterpart (canonical spin) is able to counterbalance the total SAM density of the field as necessary.
Interplay between spin's canonical and Poynting components creates counter-intuitive effects, such as spin-free chiral interaction forces. 

\subsection{Evanescent wave}
The electric and magnetic field phasors of an evanescent wave of arbitrary polarisation, propagating in the $z$ direction ($k_z > k$) and decaying along the $x$ axis with decay constant $\gamma = \sqrt{k_z^2-k^2}$, are,
\begin{equation}\label{EHevanescent}
    \mathbf{E}=\begin{pmatrix}
        A_p\frac{k_z}{k}\\A_s\\-iA_p\frac{\gamma}{k} 
    \end{pmatrix}e^{ik_zz-\gamma x} \quad
    \mathbf{H}=\frac{1}{\eta}\begin{pmatrix}
        -A_s\frac{k_z}{k}\\A_p\\iA_s\frac{\gamma}{k}
    \end{pmatrix}e^{ik_zz-\gamma x},
\end{equation}
where $\eta=\sqrt{\frac{\mu_0}{\epsilon_0}}$.
Choosing complex values for $A_s$ and $A_p$, which are the amplitudes of the evanescent wave's TE and TM modes, controls the wave's polarisation---a circularly polarised wave, for instance, has $A_p=\pm iA_s$.
The energy density of the wave decays in the $x$ direction and is given by,
\begin{equation}\label{W_evanescent}
    W=\epsilon_0\frac{k_z^2}{k^2}e^{-2\gamma x}\frac{1}{2}\left(|A_s|^2+|A_p|^2\right).
\end{equation}
Using our formulae, we can calculate the energy-normalised total spin of the evanescent wave, as well as its canonical and Poynting spins,
\begin{equation}\label{stot_evanescent}
    \frac{\mathbf{S}}{W}=\frac{1}{\omega k_z} \left[ \gamma \mathbf{\hat{y}} + k \sigma  \mathbf{\hat{z}} \right],
\end{equation}
\begin{equation}\label{scan_evanescent}
    \frac{\mathbf{s}_c}{W}=\frac{1}{\omega k_z}\left[ \frac{k_z^2}{k} \sigma \mathbf{\hat{z}} \right],
\end{equation}
\begin{equation}\label{spoy_evanescent}
    \frac{\mathbf{s}_p}{W}=\frac{1}{\omega k_z} \left[ \gamma \mathbf{\hat{y}} - \frac{\gamma^2}{k} \sigma \mathbf{\hat{z}} \right].
\end{equation}
The parameter $\sigma=2\textrm{Im}\{A_sA_p^*\}/(|A_s|^2+|A_p|^2)$ is the degree of circular polarisation in the sense of a plane wave ($\sigma=\pm1$ for circular polarisation, $\sigma=0$ for linear polarisation). 
It is well-known that evanescent waves carry transverse spin independently of polarisation \cite{Bliokh2014}.
This property is accounted for by the $\mathbf{\hat{y}}$ component of $\mathbf{S}$, which is unaffected by the relationship between $A_s$ and $A_p$ and, interestingly, is a product solely of Poynting spin as was discovered in \cite{Shi2021}.
Meanwhile, the evanescent wave acquires a longitudinal spin component if $\sigma\neq0$, a component which is contributed to by both decomposed spins, and in different amounts.
The over-generous canonical spin develops a larger $\mathbf{\hat{z}}$ component than is physical for the total spin of the wave ($\mathbf{s}_c\cdot\mathbf{\hat{z}}>\mathbf{S}\cdot\mathbf{\hat{z}}$).
Compensating, the Poynting spin's $\mathbf{\hat{z}}$ component points backwards to ensure $(\mathbf{s}_c+\mathbf{s}_p)\cdot\mathbf{\hat{z}}=\mathbf{S}\cdot\mathbf{\hat{z}}$ (this becomes clear after substituting $\gamma^2=k_z^2-k^2$).

Decomposing an evanescent wave's spin reveals a physical distinction between its transverse and longitudinal spin components.
Since Poynting spin is responsible for the transverse component of $\mathbf{S}$, transverse chiral forces felt by an enantiomer arise as it recoils from its own radiation, rather than from a direct field interaction.
This is consistent with the fact that perpendicular to the wavevector, there is no phase advance to twist the rotating field vectors into helices---the evanescent wave only acquires helicity if it carries longitudinal spin ($h=W\sigma/\omega$). 
Canonical spin, purely longitudinal on the other hand, couples directly to chiral matter to produce a significantly stronger (relatively) preferential force.

\subsection{Beams}

\begin{figure}[H]
    \centering
    \includegraphics[width=0.9\textwidth]{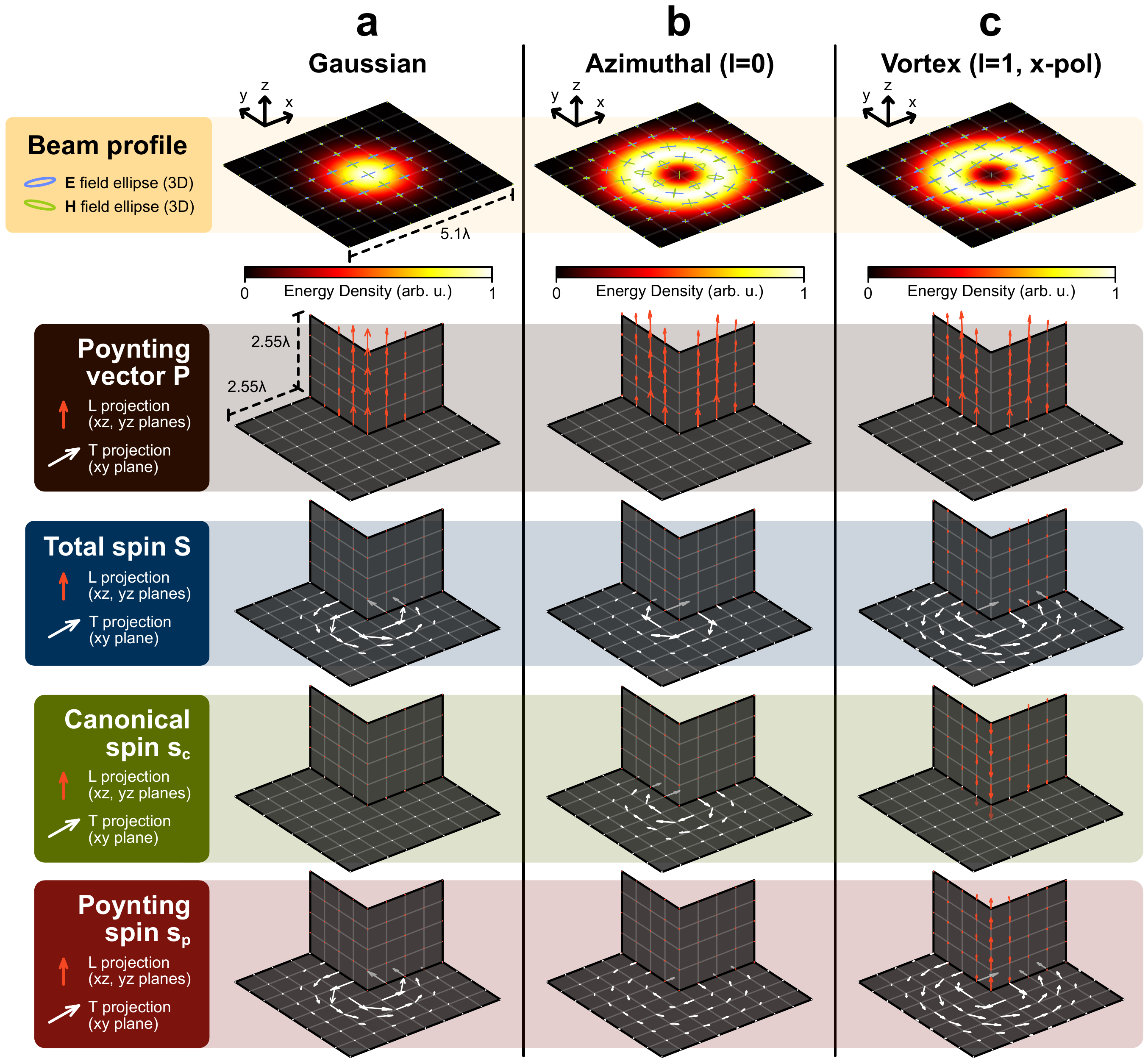}
    \caption{
    Spin decomposition of non-paraxial beams.
    The beams, each of waist $1.5\lambda$ and separated in columns, are (a) a linearly polarised Gaussian beam ($\mathbf{E}^T||\mathbf{\hat{x}}$), (b) an azimuthally polarised ($\mathbf{E}||\boldsymbol{\hat{\phi}}$), $l=0$ doughnut beam, and (c) a linearly polarised ($\mathbf{E}^T||\mathbf{\hat{x}}$) vortex beam with topological charge $l=1$.
    The top row of (isometric) plots across each sub figure column shows the beam energy density in colour, as well as the Poynting vector and electric (blue) and magnetic (green) polarisation ellipses, which are elliptical due to a significant $z$ field component.
    Subsequent rows are vector plots of each beam's Poynting vector $\mathbf{P}$, total spin $\mathbf{S}$ Eq. (\ref{def_spin}), canonical spin $\mathbf{s}_c$ and Poynting spin $\mathbf{s}_p$ from Eq. (\ref{spin_decomp}), respectively.
    White arrows are projections of the corresponding vector into the $xy$ plane, while the red arrows are projections of the vector onto longitudinal $yz$ and $xz$ cut planes.
    Within each beam, arrows in the three spin decomposition plots are drawn to a consistent scale.
    Each non-paraxial beam is generated using an angular spectrum integration method \cite{KS2023}.
    Defining the 3D vector vortex beam (c) is a difficult problem and the method we used produces a small and physical longitudinal spin (third row of (c)), contributed by the magnetic field, which would not be present in a (non-physical) perfect paraxial beam.
    }
    \label{fig1}
\end{figure}

Let us first consider a tightly focussed, linearly polarised Gaussian beam, whose energy density and polarisation in the focal plane is plotted in the top row of Fig. \ref{fig1}(a).
In the transverse $xy$ plane, the projection of the electric and magnetic field vectors are linear ($x$ and $y$ polarised respectively), though due to the tight focusing both $\mathbf{E}$ and $\mathbf{H}$ have significant longitudinal $z$ components, such that they are elliptically polarised in 3D and contribute a circulation of transverse spin.
%Meanwhile, the beam has approximately planar phase fronts such that the Poynting vector $\mathbf{P}$ is longitudinal (with a small radial component due to diffraction).
The Poynting vector magnitude is not constant across the face of the beam and therefore has a non-zero curl in the transverse plane (except in the local maximum at the beam centre).
This, the total spin and the decomposed canonical and Poynting spin of the Gaussian beam are plotted in Fig. \ref{fig1}(a)'s column (descending).
Both $\mathbf{E}$ and $\mathbf{H}$ fields have $z$ components and discretely symmetric polarisation ellipses in 3D (after a rotation of one field by $\pi/2$), which appears to reduce the canonical spin of the beam to zero, $\mathbf{s}_c=\mathbf{0}$.
All transverse spin developed by focussing of a linearly polarised Gaussian beam stems from Poynting spin, $\mathbf{S}=\mathbf{s}_p$.
A similar conclusion could be made for an evanescent wave, making it tempting to argue for a general association between transverse spin and Poynting spin.
This notion is not supported by the following example, however.

An $l=0$ azimuthally polarised (in the electric field) beam and its spin decomposition is plotted in Fig \ref{fig1}(b).
Considering dual spin (contributions from both $\mathbf{E}$ and $\mathbf{H}$), the decomposition properties of this beam are identical for a radially polarised beam, which has an azimuthal magnetic field.
Despite tight focussing, the azimuthal electric field vector does not carry a longitudinal component, as is well-known, and is completely linearly polarised in 3D.
This means that the beam's wholly transverse spin is supplied by the magnetic field alone.
Combining to give the total spin of the beam, both canonical and Poynting spins are non-zero---this contrasts with the Gaussian beam, whose electric and magnetic fields are both elliptical in 3D.
Neither beam considered so far has twisted wavefronts carrying OAM, though due to their azimuthal transverse spin components, both the Gaussian and radial beam possess a certain chiral OAM.
A chiral particle could access the spatial structure of the beam's spin field $\mathbf{S}$, feeling an azimuthal force (clockwise or anticlockwise depending on the enantiomer) which would cause the particle to orbit the beam centre with no need for helical phasefronts.

Finally, we treat the $x$-polarised $l=1$ vortex beam [Fig. \ref{fig1}(c)], whose OAM leads to the most surprising spin decomposition features of the three.
Numerically simulating such a non-paraxial, focussed beam is a challenging task because there is no unanimous 3D definition of a vector vortex beam (only scalar and paraxial beams are well-defined). 
The beams treated in this section are all electric linearly polarised and generated numerically using an angular spectrum integration technique \cite{KS2023} which, by convention, forces the electric field to exactly match the paraxial description of the beam in the focal plane.
All unenforced field components (i.e., all magnetic field components and the longitudinal electric component throughout space, and the transverse electric components outside of the focal plane) are subsequently calculated via Maxwell's equations.
%While the beams are, therefore, physically valid, some polarisation features develop in the $x$-polarised vortex beam considered here which are unexpected in the paraxial picture.
While the beams are, therefore, physically valid, some magnetic-biased polarisation features develop in the $x$-polarised vortex beam considered here.
%In particular, the electric polarisation ellipses obtain a small and rotationally asymmetric $z$-tilt such that the magnetic field gains a slight ellipticity in the transverse plane (producing a small longitudinal total spin outside of the beam centre).
In particular, although the electric field remains perfectly linear, the magnetic field alone gains a slight ellipticity in the transverse plane, contributing to a small longitudinal total spin outside of the beam centre as has been shown to exist in focussed linearly polarised beams \cite{Forbes2021_2}.
Longitudinal spin remains zero in the singularity, however.
Breaking continuous rotational symmetry via $x$ polarisation, both the vortex beam's electric and magnetic fields possess a $z$ component from focussing, in contrast to the azimuthal beam for which the electric field is perfectly linearly polarised in 3D.
The beam has helical wavefronts so that the average phase gradient (local wavevector) has an azimuthal component, also inherited by the Poynting vector (shown in the second row of Fig. \ref{fig1}(c); there is a small transverse component circulating in the $xy$ plane).
The Poynting spin $\mathbf{s}_p\propto\nabla\times\mathbf{P}$, therefore, acquires a longitudinal component (even at the centre of the beam), in outright defiance of the fact that both transverse electric and magnetic fields are nearly linearly polarised, and zero in the centre, as is visible in the lowest plot of Fig. \ref{fig1}(c).
To neutralise the longitudinal Poynting spin and suppress the total longitudinal spin, the $z$ component of canonical spin $\mathbf{s}_c$ must be non-zero too, satisfying $s_{cz}=-s_{pz}$ for $S_z=s_{cz}+s_{pz}=0$ in the centre of the beam, and $S_z=s_{cz}+s_{pz}\approx0$ elsewhere (see the plot second from bottom).
In the middle of the beam, this counteracting canonical spin points in the negative $z$ direction as a consequence of the beam's vortex handedness (right, topological charge $+1$), and would switch sign in a left-handed vortex.
Returning now to the chiral force equation Eq. \eqref{eq:chiral_force}, we can infer that the beam's non-zero canonical spin, proportional to chiral momentum, produces a longitudinal chiral force, even though near to the vortex centre the $\mathbf{E}$ and $\mathbf{H}$ fields have negligible ellipticity in the transverse plane.  
Perhaps counter-intuitively, the longitudinal chiral force is strongest in the centre of the beam where the electromagnetic field is virtually zero (due, of course, to the longitudinal curl of the Poynting vector being maximal there).
This is another example of spatially chiral light coupling to chiral matter which, even under the dipole approximation and in a linearly polarised field, can feel a discriminatory force by absorbing or scattering photons carrying OAM of a certain handedness.
Only recently has OAM-dependent chirality, able to couple even to chiral dipoles, been demonstrated in focussed beams, resulting from strong longitudinal field components \cite{Forbes2022,Green2023}.
We point out, however, that the chiral pressure force which we theorise at the centre of a linearly polarised vortex appears only to require that the Poynting vector has an azimuthal component (i.e., no requirement for longitudinal field components), and should still be present in a paraxial beam.
%We point out that this canonical spin-induced force persists in the dipole approximation where preferential absorption of photons with OAM $\pm l\hbar$ does not occur \cite{Andrews2004}---the force should therefore arise from forward or back scattering of photons.

\section{Analogies in other wave fields}
Kinetic, canonical and spin momenta analogous to the terms in Eq. (\ref{poynting_decomp}) can be identified more generally in other wave fields \cite{Bliokh2022,Bliokh2019,Bliokh2019_2}. 
This, and the ability to split the SAM density of light into two distinct terms poses another curiosity: what does a spin decomposition of another wave field look like, particularly if, unlike the electromagnetic field, its quanta are not spin-1?
Treated in this section are acoustic and gravitational waves, which both depart from light's spin-1 structure.
Despite their increased complexity, linearised theories exist for each of these fields under certain conditions.
In a perfect fluid, acoustic waves are linear and oscillate longitudinally (spin-0), while gravitational waves are tensorial in nature (spin-2) and combine linearly when their amplitudes are sufficiently low.

In formulating equivalent expressions for total SAM, canonical and Poynting spins in linearised gravity, we use the Maxwellian representation of gravity in the weak field limit \cite{Barnett2014}, i.e., waves propagating over flat spacetime at a large distance from their source.
Gravitational waves detected on earth \cite{PhysRevLett.116.061102,PhysRevLett.119.161101} arrive within this limit, and share some of light's characteristics; both fields have two polarisation degrees of freedom (in gravity these are the $h_+$ and $h_{\times}$ polarisations) and are massless.
Some of light's strangest behaviour, such as in evanescent fields, have also been predicted with additional properties in gravitational waves \cite{Golat:2019aap}.
Note that we take the expressions for the helicity and SAM density that are dual-symmetric (i.e., $\mathbf{S}=\mathbf{S}_{e}+\mathbf{S}_{m}$ rather than $\mathbf{S}=2\mathbf{S}_{e}$), the same as in the case of electromagnetism, despite the fact that in the case of gravity we have no experimental evidence favouring its physical relevance over the asymmetric version \cite{Xin2021}.
However, if we take the point of view of \cite{Cameron2012} that only the integrated helicity is a physically meaningful quantity, the two will be equivalent.

Table~\ref{summary_comparison} summarises differences and similarities between linearised acoustics, electromagnetism, and linearised gravity with a focus on the spin-related quantities.
An acoustic wave field can be described by a scalar pressure field $P$ and a vector velocity field $\mathbf{v}$ which, in linearised acoustic theory, share a Maxwell-like relation \cite{Bliokh2019,Bliokh2019_2}.
The derivation of the decomposition for the acoustic field is given in the supplementary material.
A gravitational wave can be described using a metric perturbation $h_{ij}$, which can be thought of as components of a three-by-three symmetric matrix.
If we consider only $i$-th row or column of this matrix as a vector potential, $\vc{A}^i=h^{ij}\hat{\mathbf{e}}_j$, then the Maxwellian representation of gravity can be written in vector notation; the derivation of the spin decomposition then becomes simply Eq.~\eqref{decomp_step2}. The acoustic helicity density is zero, which is a feature of spin-0 fields, but interestingly the canonical spin, which we showed in Eq.~\eqref{poynting_canonical_spin} to be related to helicity, is also zero. One can see that electromagnetic and gravitational waves are spin-1 and spin-2 respectively by taking a circularly polarised wave. The helicity density of a circularly polarised wave will be one(two) times the energy density and canonical spin will be one(two) times the canonical momentum for an electromagnetic(gravitational) wave.

%Gravity and electromagnetism share very similar property at a linearized level (see e.g \cite{Golat:2019aap} or \cite{Barnett2014} for a review). Indeed, when considering linearized gravity, i.e small gravitational perturbations on top of flat spacetime, we find that the only particle propagating is a massless spin $2$ satisfying the wave equation \cite{2004sgig.bookC}. These correspond to gravitational waves, that were detected for the first time less than a decade ago \cite{PhysRevLett.116.061102,PhysRevLett.119.161101}.  
%Just like light, gravitational waves have $2$ degree of polarizations, $h_+$ and $h_{\times}$ and are massless. 
%Therefore it is interesting to compare electromagnetism and linearized gravity to see whether or not we can describe them using similar equation. Remarkably, this was done in \cite{Barnett2014} and the conclusion is the following: it is possible to write Maxwell-like equations describing linearized gravity. The main difference is that there will be an extra index that will be internally summed up on and an overall extra factor of $2$. The extra factor of $2$ is consistent with the fact that the graviton (the massless propagator for linearized gravity) is spin $2$ whereas the photon is spin $1$ 

\begin{table}[H]
\makebox[\textwidth][c]{
\begin{tabular}{c c c c}
& \textbf{Linearised Acoustics} & \textbf{Electromagnetism} & \textbf{Linearised gravity}\\\\[-0.8em]\hline\\
            
%Potential field(s) & Scalar $\varphi$ & Scalar $\phi$, vector $\mathbf{A}$ & Tensor $h_{\mu \nu}$\\\\\hline\\

%Choice of gauge  & - & \makecell{Coulomb gauge:\\\\[-0.9em] $\phi=0$ \\\\[-0.9em] $\mathbf{\nabla}\cdot \mathbf{A}=0$} & \makecell{TT gauge:\\\\[-0.9em] $h_{0 \mu}^{\mathrm{TT}}=0$ \\\\[-0.9em] $h_\mu^{\mathrm{TT} \mu}=0=h_{i i}^{\mathrm{TT}}$ \\\\[-0.9em] $h_{i j, i}^{\mathrm{TT}}=0=h_{j i, i}^{\mathrm{TT}}$} \\\\\hline\\

%Wave equation & $\Box\, \varphi =0$ & $\Box\, \mathbf{A} =0$ & $\Box\, h_{ij}^{TT}=0$ \\\\\hline\\

Field phasors & \makecell{$P=-i\omega\rho\varphi$ \\\\[-0.9em] $\mathbf{v}=\mathbf{\nabla} \varphi$} & \makecell{$\mathbf{E}=i\omega\mathbf{A}$ \\\\[-0.9em] $\mathbf{H}=\frac{1}{\mu_0}\mathbf{\nabla}\times\mathbf{A}$} & \makecell{$\mathbf{E}^i=i\omega (h^{ij}\hat{\mathbf{e}}_j)$ \\\\[-0.9em] $ \mathbf{H}^{i}=\frac{1}{\mu_0}\nabla\times(h^{ij}\hat{\mathbf{e}}_j)$}
\\\\\hline\\

Energy density &$\frac{1}{4} \!\left(\beta |P|^2+\rho |\mathbf{v}|^2\right)$ & $\frac{1}{4}\!\left(\epsilon_0|\mathbf{E}|^2+\mu_0|\mathbf{H}|^2\right)$ & $\frac{1}{4}\!\left(\epsilon_0\mathbf{E}^*_i\cdot\mathbf{E}^i+\mu_0\mathbf{H}^*_i\cdot\mathbf{H}^i\right)$\\\\\hline\\

Helicity density & $0$ & $-\frac{1}{2\omega c}\textrm{Im}\{\mathbf{E^*\cdot H}\}$ & $-\frac{1}{\omega c}\textrm{Im}\{\mathbf{E}_i^*\cdot \mathbf{H}^i\}$\\\\\hline\\

Poynting vector& $\frac{1}{2}\textrm{Re}\{P^*\mathbf{v}\}$ & $\frac{1}{2}\textrm{Re}\{\mathbf{E^*\times H}\}$ & $\frac{1}{2}\textrm{Re}\{\mathbf{E}_i^*\times \mathbf{H}^i\}$ \\\\\hline\\

SAM density &  $\frac{1}{2\omega}\textrm{Im}\{\rho\mathbf{v^*}\times\mathbf{v}\}$ & $\frac{1}{4\omega}\textrm{Im}\{\epsilon_0\mathbf{E^*}\times\mathbf{E}+\mu_0\mathbf{H^*}\times\mathbf{H} \}$ & $\frac{1}{2\omega}\textrm{Im}\{\epsilon_0\mathbf{E}^*_i\times\mathbf{E}^i+\mu_0\mathbf{H}^*_i\times\mathbf{H}^i \}$ \\\\\hline\\

Canonical spin & $0$ & $\frac{1}{4\omega^2}\textrm{Re}\{\mathbf{E^*}\cdot(\nabla)\mathbf{H}-\mathbf{H^*}\cdot(\nabla)\mathbf{E}\}$ & $\frac{1}{2\omega^2}\textrm{Re}\{\mathbf{E}^*_i\cdot(\nabla)\mathbf{H}^i-\mathbf{H}^*_i\cdot(\nabla)\mathbf{E}^i\}$ \\\\\hline\\

Poynting spin & $\frac{1}{2\omega^2}\nabla\times\frac{1}{2}\textrm{Re}\{P^*\mathbf{v}\}$ & $\frac{1}{2\omega^2}\nabla\times\frac{1}{2}\textrm{Re}\{\mathbf{E^*\times H}\}$ & $\frac{1}{\omega^2}\nabla\times\frac{1}{2}\textrm{Re}\{\mathbf{E}^*_i\times \mathbf{H}^i\}$ \\\\\hline

%\makecell{Maxwell(-like)\\equations} & \makecell{$\mathbf{\nabla} P=i \rho \omega \mathbf{v}$ \\\\[-0.9em] $\mathbf{\nabla}\cdot \mathbf{v}=i \beta \omega P$ \\\\[-0.9em] $\nabla\times\mathbf{v}=\mathbf{0}$}
%     & \makecell{$\mathbf{\nabla}\cdot \mathbf{E}=0$ \\\\[-0.9em] $\mathbf{\nabla}\cdot \mathbf{H}=0$ \\\\[-0.9em] $\mathbf{\nabla} \times \mathbf{E}=i\omega\mu_0\mathbf{H}$ \\\\[-0.9em] $\mathbf{\nabla} \times \mathbf{H}=  -i\omega\epsilon_0\mathbf{E}$}
%     & \makecell{$\nabla_i E^{ij}=0$ \\\\[-0.9em] $\nabla_i H^{ij}=0$ \\\\[-0.9em] $\varepsilon^{ijk}\nabla_j  E_{km}=i\omega H^{im}$ \\\\[-0.9em] $\varepsilon^{ijk}\nabla_j  H_{km}=-i\omega\frac{1}{c^2}E^{im}$}\\\\\hline

\end{tabular}
}
\caption{
Comparison of time-averaged energy, momentum and spin densities between time-harmonic waves in theories of linearised acoustics, electromagnetism and linearised gravity. 
Table inspired by \cite{Bliokh2019,Bliokh2022,Golat:2019aap}. For electromagnetism, the potential is considered to be in the Coulomb gauge. For linearised gravity, $h_{ij}$ are spatial components of the metric perturbation in the
transverse-traceless gauge, $\hat{\mathbf{e}}_i$ are basis vectors, and any repeated indices are summed over (Einstein's convention). Parameters $\epsilon_0=1/(c^2 \mu_0)=c^2/(32 \pi G)$ were chosen such that the time-averaged energy density takes the same form as the expression for the electromagnetic field. 
A larger version of this table using the index notation introduced in Sec.~\ref{sec:fourvector} is given in the supplementary material.
}
\label{summary_comparison}
\end{table}

\section{Conclusions}

We have discussed a known but overlooked decomposition which exists for the spin angular momentum density of light, in a similar way to the Poynting vector, which can be split into orbital and spin currents.
Spin is decomposed into two terms, which we call the canonical spin $\mathbf{s}_c$ and Poynting spin $\mathbf{s}_p$.
We have further expressed the decomposition for time-varying fields, as well as in four-vector notation.
The canonical spin is proportional to chiral linear momentum which can be transferred to chiral enantiomers in a positive or negative direction, depending on the enantionmer handedness.
The resulting force is a chiral analogy to the (achiral) force due to radiation pressure---both are relatively strong first order forces and can act on matter in the dipole approximation.

The mechanism for preferential photon absorption or scattering by chiral matter is normally associated with the photon's spin state (i.e., circular dichroism), and less so to the sign of its OAM (although these interactions have received recent attention).
We emphasised, however, that light is capable of exerting a first-order chiral force which does not directly depend on the total SAM of light, rather, its canonical spin, which can take a non-zero value even in linearly polarised fields.
This is due to the second term in the spin decomposition, the Poynting spin, which by definition (being the curl of the Poynting vector) depends strongly on light's OAM: optical vortices twisting energy flow around the beam axis naturally generate non-zero Poynting spin, regardless of polarisation.
We showed that under linear polarisation, an optical vortex's Poynting spin must be compensated by canonical spin, to produce a longitudinal chiral force in the absence of longitudinal SAM.
Interestingly, this OAM-dependent chiral force is strongest in the centre of the vortex, where the electromagnetic energy density is minimal, along with achiral forces such as radiation pressure and gradient force.
It is possible to trap atoms and molecules (using a blue-detuned wavelength with respect to a strong resonance) in dark spots \cite{He1995,Grimm2000,Vetsch2010}.

The ability to decompose spin appears to be a general property of wave fields, and as we demonstrated, is straightforward to perform in linearised acoustics and gravity, whose quanta are considered spin-0 and spin-2 respectively.
This fact could be of significant interest to a broader community, beyond optics.

\section{Acknowledgements}
C. R.  acknowledges support from a Science and Technology Facilities Council (STFC) Doctoral Training Grant. 
F. J. R-F. and S. G. are supported by EIC-Pathfinder-CHIRALFORCE (101046961) funded by Innovate UK Horizon Europe Guarantee (UKRI project 10045438).
A. J. V. is supported by EPSRC Grant EP/R513064/1.

\clearpage

\section{Supplementary Information}

This section supports the main text with four main components.
In section \ref{alttable}, we provide an alternative version of Table I of the manuscript, expressed completely using index notation, which highlights the similarities and differences between vector and tensor waves.
Secondly, section \ref{acoustics} is a discussion of the equivalent SAM density decomposition in linerised acoustics, which informs the first column of Table I of the manuscript.
In section \ref{units}, details of the units of the quantities handled in the manuscript's 4-vector decomposition are given.
Finally in section \ref{gauge}, the electromagnetic spin decomposition is presented in a general gauge, supporting comments made in the in 4-vector section of the main text.

%\clearpage

\subsection{Alternative table (index notation)}\label{alttable}

An alternative version of Table 1 from the manuscript is given below, where all quantities are expressed using index notation.

\begin{table}[H]
\makebox[\textwidth][c]{
\begin{tabular}{c c c c}

& \textbf{Linearised acoustics} & \textbf{Electromagnetism} & \textbf{Linearised gravity} \\\\[-0.8em]\hline\\

Potential field & Scalar field $\varphi$ & Vector field $A^\mu$ & Tensor field $h_{\mu \nu}$\\\\\hline\\

\makecell{Choice of\\gauge}  & n/a & \makecell{Coulomb gauge:\\$A^0=0$\\$\partial_iA^i=0$}  & \makecell{Transverse traceless gauge:\\$h_{0\mu}=h^{i}{}_i=0$\\$\partial_ih^{ij}=0$}\\\\\hline\\

\makecell{Helmholtz\\equation} & $\nabla^2 \varphi =-k^2\varphi$ &   $\nabla^2 A^i =-k^2A^i$    &   $ \nabla^2 h^{ij}=-k^2h^{ij}$\\\\\hline\\

\makecell{Fields}& \makecell{$P=-i\omega\rho \varphi$ \\ $v^i= \partial^i \varphi$} & \makecell{$E^i=i\omega A^i$\\$\mu_0 H^i= \epsilon^{ijk}\partial_j A_k$} & \makecell{ $E^{ij}=i\omega h^{ij}$\\$\mu_0 H^{ij}= \epsilon^{ikm}\partial_k h^j_{~m}$}\\\\\hline\\

Spin & $\frac{\rho}{2\omega}\epsilon^{ijk}\textrm{Im}\{v^{*}_jv_k\}$  & $\frac{1}{4\omega}\epsilon^{ijk}\textrm{Im}\{\epsilon_0 E^*_j  E_k+\mu_0 H^*_j H_k\}$ & $\frac{1}{2\omega}\epsilon^{ijk}\textrm{Im}\{\epsilon_0 E^*_{jl}  E_{k}{}^l+\mu_0 H^*_{jl} H_{k}{}^l\}$\\\\\hline\\

Canonical spin & $0$ & $\frac{1}{4\omega^2}\textrm{Re}\{E^*_j \partial_i H^j-H^*_j \partial_i E^j\}$ & $\frac{1}{2\omega^2}\textrm{Re}\{E^*_{jn}\partial_iH^{jn}-H^*_{jn}\partial_iE^{jn}\}$\\\\\hline\\

Poynting spin & $\frac{1}{2\omega^2} \epsilon^{ijk}\partial_j \frac{1}{2}\textrm{Re}\{P^*v_k\}$ & $\frac{1}{2\omega^2} \epsilon^{ijk}\partial_j \frac{1}{2}\textrm{Re}\{\epsilon_{klm} E^{*l} H^m\}$ & $\frac{1}{\omega^2}\varepsilon^{ijk}\partial_j\frac{1}{2}\textrm{Re}\{\varepsilon_{klm}E^{*l}{}_{n}H^{mn}\}$\\\\\hline\\

Energy density & $\frac{1}{4} \{ \beta p^*p+\rho v_i^*v^i\}$ & $\frac{1}{4} \{\epsilon_0 E^*_j  E^j +\mu_0 H^*_j H^j\}$ & $\frac{1}{4} \{\epsilon_0 E^*_{jk}  E^{jk} +\mu_0 H^*_{jk} H^{jk}\}$\\\\\hline\\

\makecell{Maxwell(-like)\\equations} & \makecell{$\partial_i P=i\omega \rho  v^i$\\\\[-1em]$\partial_i v^i=i\omega \beta  P$\\\\[-1em]$\epsilon^{ijk}\partial_j  v_k=  0$} & \makecell{$\partial_i E^i=0$\\\\[-1em]$\partial_i H^i=0$\\\\[-1em]$\epsilon^{ijk}\partial_j  E_k=i\omega \mu_0H^i$\\\\[-1em]$\epsilon^{ijk}\partial_j  H_k=  -i\omega \epsilon_0 E^i$} & \makecell{$\partial_i E^{ij}=0$\\\\[-1em]$\partial_i H^{ij}=0$\\\\[-1em]$\epsilon^{ijk}\partial_j  E_{k}{}^l=i\omega \mu_0H^{il}$\\\\[-1em]$\epsilon^{ijk}\partial_j  H_{k}{}^l=  -i\omega \epsilon_0 E^{il}$}\\\\\hline\\

\end{tabular}
\label{summary_comparison_2}
}
\caption{
Comparison between acoustics, electromagnetism and linearised gravity. Table inspired by \cite{Bliokh2019,Bliokh2022,Golat:2019aap}.
The parameter $\epsilon_0=1/(c^2 \mu_0)=c^2/(32 \pi G)$ for linearised gravity was chosen such that the time-averaged energy density takes the same form as for electromagnetism \cite{Barnett2014}.
}
\end{table}

\subsection{Decomposition in Linearised acoustics}\label{acoustics}

An acoustic wave field can be described by a scalar pressure field $P$ and a vector velocity field $\mathbf{v}$ which, in linearised acoustic theory, share a Maxwell-like relation \cite{Bliokh2019,Bliokh2019_2}.
The time-harmonic equations are,
\begin{equation}\label{divv}
    \nabla\cdot\mathbf{v}=i\beta\omega P,
\end{equation}
\begin{equation}\label{gradP}
    \nabla P=i\rho\omega\mathbf{v}.
\end{equation}
where the constants $\beta$ and $\rho$ are the acoustic medium's compressibility and mass density respectively.
Compared to photons, acoustic phonons, which are spin-0 quanta in this regime, do not give acoustic fields as rich a vector structure as light.
Constraining the velocity vector is the longidutinality condition, that is $\nabla\times\mathbf{v}=\mathbf{0}$, a more restrictive analogy to light's transversality condition due to Gauss' law.
Yet, $\mathbf{v}$ can still rotate and generate acoustic SAM, which is expressed (time-averaged) by,
\begin{equation}
    \mathbf{S}_{ac}=\frac{\rho}{2\omega}\textrm{Im}\{\mathbf{v^*\times v}\}.
\end{equation}
In time-harmonic acoustic fields, where the velocity $\mathbf{v}$ and displacement field $\mathbf{r}$ vectors are related by $\mathbf{v}=-i\omega\mathbf{r}$, particles in the acoustic medium have an elliptical motion.
Meanwhile, an acoustic analogy to the Poynting vector can be defined by mixing the acoustic pressure and velocity fields,
\begin{equation}
    \mathbf{P}_{ac}=\frac{1}{2}\textrm{Re}\{P^*\mathbf{v}\}.
\end{equation}
Taking the curl of $\mathbf{P}_{ac}$ and following with use of Eq. (\ref{gradP}), the acoustic SAM emerges in what appears to be the acoustic analogy to the presented electromagnetic spin decomposition:
\begin{equation}\label{acdecomp}
    \mathbf{S}_{ac}=-\frac{1}{2\omega^2}\textrm{Re}\{P^*(\nabla\times\mathbf{v})\}+\frac{1}{2\omega^2}\nabla\times\mathbf{P}_{ac}.
\end{equation}
Due to the longitudinality condition $\nabla\times\mathbf{v=0}$, the first term in Eq. (\ref{acdecomp}) vanishes leaving the acoustic spin entirely proportional to the curl of the acoustic Poynting vector.
Transverse phonons for which $\nabla\times\mathbf{v\neq0}$ can occur in viscous fluids or solids, although in these media, the acoustic field could no longer be described by a linearised theory.
Compared to the electromagnetic spin decompostion, however, Eq. (\ref{acdecomp}) highlights the structural distinction between spin-0 and spin-1 fields as only vorticity in the flow of energy can generate SAM in a linear acoustic field.

\subsection{On units}\label{units}
Discussions in this section are related to the 4-vector extension of the decomposition presented in section 4.2 of the manuscript.
The field strength tensor $F_{\mu \nu}$ and conjugate field strength tensor $G_{\mu \nu}$ are not of the same units.
The former has SI units \si{kg.s^{-4}.A^{-1}} whilst the latter has SI units \si{m^{-1}.kg.s^{-5}.A^{-1}}.
This comes from the form of the conjugate field strength tensor that can be derived using the companion Mathematica notebook \cite{Golat_Rigouzzo}: 
 \begin{equation}
G_{\mu \nu}=\left[\begin{array}{cccc}
0 &  H_x /c & H_y /c  & H_z /c \\
-H_x/c & 0 & -  \epsilon_0 E_z & \epsilon_0 E_y \\
-H_y /c  & \epsilon_0 E_z & 0 &  - \epsilon_0 E_x \\
-H_z /c & -  \epsilon_0E_y &  H_x /c^2 & 0
\end{array}\right].
\end{equation}
Finally, we summarize all the SI units of the mentioned tensor and constants: 
\begin{table}[h]
\begin{tabular}{|c|c|c|c|c|c|c|c|}
\hline
         & \textbf{E} & \textbf{H} & $\epsilon_0$ & $\mu_0$ & \textbf{A} & $\phi$ & \textbf{C} \\ \hline 
 Units &    \si{m.kg.s^{-3}.A^{-1}}  &  \si{A.m^{-1}} &   \si{kg^{-1}.m^{-3}.s^4.A^{-2}}           &  \si{kg.m.s^{-2}.A^{-2}}       &     \si{kg. m.s^{-2}. A^{-1}}                        &  \si{kg . m^2. s^{-3}. A^{-1}}      &    \si{A.m^{-1}.s}                         \\ \hline
\end{tabular}
\caption{Summary of the SI units of every tensors and constants mentioned in this work}
\end{table}

\subsection{On the choice of gauge}\label{gauge}
\label{app_gauge}
The electric field and magnetic field, respectively $\mathbf{E}$ and $\mathbf{H}$, can be expressed in terms of a scalar potential $\phi$ and a vector potential $\textbf{A}$,
\begin{equation}
    \begin{split}
        &\boldsymbol{\mathcal{E}}=-\frac{\partial \boldsymbol{\mathcal{A}}}{ \partial t}- \mathbf{\nabla} \varphi \\
        & \mu_0 \boldsymbol{\mathcal{H}}=\mathbf{\nabla} \times \boldsymbol{\mathcal{A}}
    \end{split}
\end{equation} and respectively for the conjugate vector potential: 
\begin{equation}
    \begin{split}
         \epsilon_0\boldsymbol{\mathcal{E}}&=- \mathbf{\nabla} \times \boldsymbol{\mathcal{C}} \\
         \boldsymbol{\mathcal{H}}&=-\frac{\partial \boldsymbol{\mathcal{C}}}{\partial t}-\mathbf{\nabla}\psi
    \end{split}
\end{equation}
When doing computations in the Coulomb gauge, where the scalar potentials $\varphi=\psi=0$, we can easily express the electric and magnetic field in terms of the vector potentials through:
\begin{equation}
     E^j= i \omega A^j(\omega,r), \qquad H^j=i \omega C^j(\omega,r),
 \end{equation}
 which implies that we can express the four vector potential as: 
 \begin{equation}
A^\mu= \frac{1}{i \omega} \begin{pmatrix}
0 \\
\mathbf{E}
    \end{pmatrix}, \qquad C^\mu= \frac{1}{i \omega} \begin{pmatrix}
0 \\
\mathbf{H}
    \end{pmatrix}.
 \end{equation}
Working in the Coulomb gauge simplifies greatly the form of the spin decomposition, giving a clearer physics intuition. However, it is possible to derive all the results presented in the main work without choosing a gauge. In a general gauge, we found that the total spin is given by:
 \begin{equation}\label{eq:spin_density_vector}
\begin{split}
    S^\mu&= \frac{1}{4} \operatorname{Re}\left\{A_\nu^* G^{\nu \mu}+ C_\nu^* F^{\nu \mu}\right\}
    =
    \frac{1}{4}\operatorname{Re}\left\{\left(
    \begin{array}{c}
\frac{1}{c}\left(\mathbf{A} \cdot \mathbf{H}^*-\mathbf{C} \cdot \mathbf{E}^*\right)\\
\mu \mathbf{H}^* \times \mathbf{C}+\varepsilon \mathbf{E}^* \times \mathbf{A}
\end{array}
    \right)+\frac{1}{c^2}\left(\begin{array}{c}
0\\
\phi\mathbf{H}^*-\psi\mathbf{E}^*
\end{array}\right)\right\}\\
% &=
%     \frac{1}{4\omega}\operatorname{Im}\left\{\left(
%     \begin{array}{c}
% \frac{1}{c}\left(\mathbf{E} \cdot \mathbf{H}^*-\mathbf{H} \cdot \mathbf{E}^*\right)\\
% \mu \mathbf{H}^* \times \mathbf{H}+\varepsilon \mathbf{E}^* \times \mathbf{E}
% \end{array}
%     \right)+\left(\begin{array}{c}
% \frac{1}{c}{\nabla} \cdot\left(\psi \mathbf{E}^*-\varphi \mathbf{H}^*\right) \\
% {\nabla} \times\left(\varepsilon \varphi \mathbf{E}^*+\mu \psi \mathbf{H}^*\right)
% \end{array}\right)\right\}\\
&=\frac{1}{4 \omega } \operatorname{Im}\left\{\left(\begin{array}{c}
 -\frac{2}{c}\mathbf{E}^* \cdot \mathbf{H} \\
 \epsilon_0\mathbf{E}^*\times \mathbf{E}+\mu_0 \mathbf{H}^* \times \mathbf{H}
 \end{array}\right)+
    \left(\begin{array}{c}
\frac{1}{c}{\nabla} \cdot\left(\psi \mathbf{E}^*-\phi \mathbf{H}^*\right) \\
{\nabla} \times\left(\varepsilon \phi \mathbf{E}^*+\mu \psi \mathbf{H}^*\right)
\end{array}\right)\right\} .
\end{split}
\end{equation}
Interestingly, the extra terms for a general gauge are total derivatives, meaning that the integrated helicity is gauge invariant up to a boundary term.
This is because generically, total derivatives vanish under suitable boundary conditions. 
The canonical and Poynting spin now read as: 
% \begin{equation}
% \begin{aligned}
% &  S_{C}^\mu=\frac{1}{4} \operatorname{Re}\left\{A_\nu^*\left(\partial^\mu C^\nu\right)-C_\nu^*\left(\partial^\mu A^\nu\right)\right\}=\frac{1}{4 \omega^2}\left(\begin{array}{c}
% \text{Im}[\frac{1}{c}(\mathbf{A}\cdot \mathbf{C^*}-\mathbf{C}\cdot \mathbf{A^*})+\frac{\omega^3}{c^3}(\psi \phi^*-\phi \psi^*)] \\
% \operatorname{Re}[\omega^2(\psi^*\mathbf{\nabla}\phi-\phi^*\mathbf{\nabla}\psi)+F(\mathbf{A},\mathbf{C},\phi,\psi)]
% \end{array}\right), \\
% &  S_{P}^\mu=\frac{1}{4} \operatorname{Re}\left\{C_\nu^*\left(\partial^\nu A^\mu\right)-A_\nu^*\left(\partial^\nu C^\mu\right)\right\}=\frac{1}{4 \omega^2}\left(\begin{array}{c}
% -\text{Im}[\frac{\omega}{c}(\mathbf{C^*}\cdot \mathbf{\nabla}\phi-\mathbf{A^*}\cdot \mathbf{\nabla}\psi)+\frac{\omega^3}{c^3}(\psi \phi^*-\phi \psi^*)]\\
%  \operatorname{Re}[\frac{\omega^2}{c^2}(\mathbf{C}\cdot \phi^*-\mathbf{A}\cdot \psi^*)-F(\mathbf{A},\mathbf{C},\phi,\psi)]
% \end{array}\right). \\
% & \label{poynting_canonical_spin}
% \end{aligned}
% \end{equation}
% where $F(\mathbf{A},\mathbf{C},\phi,\psi)=\mathbf{A^*}\cdot \{(\mathbf{\nabla})\mathbf{H}+(\mathbf{\nabla})\mathbf{\nabla}\psi\}+\mathbf{C^*}\cdot \{(\mathbf{\nabla})\mathbf{E}+(\mathbf{\nabla})\mathbf{\nabla}\phi\}$.
\begin{equation}
\begin{aligned}
&  S_{C}^\mu=\frac{1}{4} \operatorname{Re}\left\{A_\nu^*\left(\partial^\mu C^\nu\right)-C_\nu^*\left(\partial^\mu A^\nu\right)\right\}=\frac{1}{4}\operatorname{Re}\left\{
\begin{pmatrix}
    -{2i\omega}(\phi^*\psi)/c^3+{i\omega}(\mathbf{A}^*\cdot\mathbf{C}-\mathbf{C}^*\cdot\mathbf{A})/c
    \\
    
    \mathbf{A}^*\cdot(\nabla)\mathbf{C}-\mathbf{C}^*\cdot(\nabla)\mathbf{A}+
    \frac{1}{c^2}(\psi^*\nabla\phi-\phi^*\nabla\psi)
    
\end{pmatrix}\right\}
, \\
&  S_{P}^\mu=\frac{1}{4} \operatorname{Re}\left\{C_\nu^*\left(\partial^\nu A^\mu\right)-A_\nu^*\left(\partial^\nu C^\mu\right)\right\}=\frac{1}{4}\operatorname{Re}\left\{\begin{pmatrix}
    {2i\omega}(\phi^*\psi)/c^3
    +\left[
    (\mathbf{C}\cdot\nabla)^*\phi-(\mathbf{A}\cdot\nabla)^*\psi
    \right]/c
    \\
    
    \nabla\times(\mathbf{A}\times \mathbf{C}^*)-\mathbf{A}^*(\partial_\mu C^\mu)+\mathbf{C}^*(\partial_\mu A^\mu)
    
\end{pmatrix}\right\} , \label{poynting_canonical_spin}
\end{aligned}
\end{equation}
where $\partial_\mu A^\mu=-i\omega \phi/c^2+\nabla\cdot\mathbf{A}$ and same for $C$, which is quantity that is zero in both Lorentz and Coulomb gauge.
An instantaneous dual symmetric gauge independent helicity current density four-vector is \cite{Cameron_2012}
\begin{equation}\label{eq:spin_density_vector_instant}
\begin{split}
    S^\mu&= \frac{1}{2}\left(\mathcal{A}_\nu \mathcal{G}^{\nu \mu}+ \mathcal{C}_\nu \mathcal{F}^{\nu \mu}\right)
    =
    \frac{1}{2}\left(
    \begin{array}{c}
\frac{1}{c}\left(\boldsymbol{\mathcal{A}} \cdot \boldsymbol{\mathcal{H}}-\boldsymbol{\mathcal{C}} \cdot \boldsymbol{\mathcal{E}}\right)\\
\mu \boldsymbol{\mathcal{H}} \times \boldsymbol{\mathcal{C}}+\varepsilon \boldsymbol{\mathcal{E}} \times \boldsymbol{\mathcal{A}}
\end{array}
    \right)+\frac{1}{2c^2}\left(\begin{array}{c}
0\\
\varphi\boldsymbol{\mathcal{H}}-\psi\boldsymbol{\mathcal{E}}
\end{array}\right).
\end{split}
\end{equation}
which can be decomposed into the canonical and Poynting spins in an arbitrary gauge as follows:
\begin{equation}
\begin{aligned}
&  S_{C}^\mu=\frac{1}{2} [\mathcal{A}_\nu\left(\partial^\mu \mathcal{C}^\nu\right)-\mathcal{C}_\nu\left(\partial^\mu \mathcal{A}^\nu\right)]=\frac{1}{2}
\begin{pmatrix}
    \frac{1}{c}\left(\boldsymbol{\mathcal{A}} \cdot \boldsymbol{\mathcal{H}}-\boldsymbol{\mathcal{C}} \cdot \boldsymbol{\mathcal{E}}\right)
    \\
    \boldsymbol{\mathcal{A}}\cdot(\nabla)\boldsymbol{\mathcal{C}}-\boldsymbol{\mathcal{C}}\cdot(\nabla)\boldsymbol{\mathcal{A}}
\end{pmatrix}
+\frac{1}{2c}
\begin{pmatrix}
    {\mathcal{A}}^\mu\partial_\mu\psi-\mathcal{C}^\mu\partial_\mu\varphi
    \\
    \frac{1}{c}\left(\psi\nabla\varphi-\varphi\nabla\psi\right)
\end{pmatrix}
, \\
&  S_{P}^\mu=\frac{1}{2} [\mathcal{C}_\nu\left(\partial^\nu \mathcal{A}^\mu\right)-\mathcal{A}_\nu\left(\partial^\nu \mathcal{C}^\mu\right)]=\frac{1}{2}\begin{pmatrix}
    0
    \\
    (\boldsymbol{\mathcal{C}}\cdot\nabla)\boldsymbol{\mathcal{A}}-(\boldsymbol{\mathcal{A}}\cdot\nabla)\boldsymbol{\mathcal{C}}
    % \nabla\times(\boldsymbol{\mathcal{A}}\times\boldsymbol{\mathcal{C}})
    
\end{pmatrix} 
-\frac{1}{2c}
\begin{pmatrix}
    {\mathcal{A}}^\mu\partial_\mu\psi-\mathcal{C}^\mu\partial_\mu\varphi
    \\
    \frac{1}{c}\left(\psi\nabla\varphi-\varphi\nabla\psi\right)
\end{pmatrix}
+\frac{1}{2c^2}\begin{pmatrix}
0\\
\varphi\boldsymbol{\mathcal{H}}-\psi\boldsymbol{\mathcal{E}}
\end{pmatrix}
, \label{poynting_canonical_spin_instant}
\end{aligned}
\end{equation}
Note that $(\boldsymbol{\mathcal{C}}\cdot\nabla)\boldsymbol{\mathcal{A}}-(\boldsymbol{\mathcal{A}}\cdot\nabla)\boldsymbol{\mathcal{C}}=\nabla\times(\boldsymbol{\mathcal{A}}\times\boldsymbol{\mathcal{C}})+\boldsymbol{\mathcal{C}}(\nabla\cdot\boldsymbol{\mathcal{A}})-\boldsymbol{\mathcal{A}}(\nabla\cdot\boldsymbol{\mathcal{C}})$ and ${\mathcal{A}}^\mu\partial_\mu\psi=\varphi(\partial\psi/\partial t)/c^2+\boldsymbol{\mathcal{A}}\cdot\nabla\psi$ and similarly  $\mathcal{C}^\mu\partial_\mu\varphi=\psi(\partial\varphi/\partial t)/c^2+\boldsymbol{\mathcal{C}}\cdot\nabla\varphi$.

\printbibliography

\end{document}